\documentclass[12pt]{article}
\usepackage{a4wide}
\usepackage{amssymb}
\usepackage{amsmath}
\usepackage{graphicx}
\usepackage{mathdots}
\usepackage{youngtab}
\usepackage{caption}
\usepackage{subcaption}
\usepackage{enumerate}

\def\drawbox#1#2{\hrule height#2pt
        \hbox{\vrule width#2pt height#1pt \kern#1pt
              \vrule width#2pt}
              \hrule height#2pt}

\def\Fund#1#2{\vcenter{\vbox{\drawbox{#1}{#2}}}}
\def\Asym#1#2{\vcenter{\vbox{\drawbox{#1}{#2}
              \kern-#2pt       
              \drawbox{#1}{#2}}}}

\def\funda{\Fund{6.5}{0.4}}
\def\asymm{\Asym{6.5}{0.4}}

\def\symm{\funda\kern-0.4pt\funda}

\def\makeatletter{\catcode`\@=11}
\makeatletter
\def\mathbox#1{\hbox{$\m@th#1$}}%
\def\math@ccstyles#1#2#3#4#5#6#7{{\leavevmode
      \setbox0\mathbox{#6#7}%
      \setbox2\mathbox{#4#5}%
      \dimen@ #3%
      \baselineskip\z@\lineskiplimit#1\lineskip\z@
      \vbox{\ialign{##\crcr
             \hfil \kern #2\box2 \hfil\crcr
             \noalign{\kern\dimen@}%
             \hfil\box0\hfil\crcr}}}}
\def\mathaccstyles{\math@ccstyles\maxdimen}
\def\maththroughstyles{\math@ccstyles{-\maxdimen}}
\def\unity%
 {\maththroughstyles{.45\ht0}\z@\displaystyle {\mathchar"006C}\displaystyle 1}

\begin{document}

\setcounter{table}{0}

\mbox{}
\vspace{2truecm}
\linespread{1.1}

\centerline{\LARGE \bf 5d superconformal indices at large $N$ and holography}

\vspace{2truecm}

\centerline{
    {\large \bf Oren Bergman ${}^{a}$} \footnote{bergman@physics.technion.ac.il},
    {\large \bf Diego Rodr\'{\i}guez-G\'omez${}^{b}$} \footnote{d.rodriguez.gomez@uniovi.es}
    {\bf and}
    {\large \bf Gabi Zafrir${}^{a}$} \footnote{gabizaf@techunix.technion.ac.il}}

\vspace{1cm}
\centerline{{\it ${}^a$ Department of Physics, Technion, Israel Institute of Technology}} \centerline{{\it Haifa, 32000, Israel}}
\vspace{1cm}
\centerline{{\it ${}^b$ Department of Physics, Universidad de Oviedo}} \centerline{{\it Avda. Calvo Sotelo 18, 33007, Oviedo, Spain }}
\vspace{1cm}

\centerline{\bf ABSTRACT}
\vspace{1truecm}

We propose a general formula for the perturbative large $N$ superconformal index of 5d
quiver fixed point theories that have an $AdS_6\times S^4/\mathbb{Z}_n$ supergravity dual.
This index is obtained from the parent theory by projecting to orbifold-invariant states and adding
the twisted sector contributions. Our result agrees with expectations from the dual supergravity
description. We test our formula against the direct computation of the index for $\mathbb{Z}_2$
and $\mathbb{Z}_3$ and find complete agreement.

\noindent

\newpage
\setcounter{footnote}{0}

\tableofcontents

\section{Introduction}

There has recently been a renewed interest in supersymmetric gauge theories in five dimensions.
On the one hand, these are strategically positioned between
the familiar four-dimensional supersymmetric gauge theories, about which we know a lot, and six-dimensional superconformal theories,
about which we know very little.
Therefore 5d theories might potentially incorporate features of the latter while allowing for the use of the well understood
techniques of the former.
On the other hand, contrary to the naive expectation, five dimensional gauge theories can be at fixed points
\cite{Seiberg:1996bd,Intriligator:1997pq}. These fixed points can have exotic properties such as $E$-type quantum-mechanically enhanced global symmetries at the origin of the Coulomb branch, and are thus very interesting \textit{per se}.

In one specific class of theories, a dual $AdS_6$ geometry was found by engineering the 5d gauge theories
using D4-branes in Type I' string theory \cite{Brandhuber:1999np}.
In \cite{Bergman:2012kr}, two of us argued for the existence of additional 5d fixed points associated with supersymmetric quiver gauge theories, obtained by orbifolding the Type I' D4-brane configuration.
One again finds $AdS_6$ near-horizon geometries, which strongly suggests that the dual quiver gauge theories are at fixed points.
A number of non-trivial checks of these new AdS/CFT proposals have recently been carried out in \cite{Bergman:2012qh,Jafferis:2012iv, Assel:2012nf}. We should point out that 6d $AdS$ backgrounds are quite rare compared with other dimensions. A no-go theorem was in fact proposed in \cite{Passias:2012vp}. Although, by allowing for more general spaces
that arise from non-abelian T-duality transformations, a new $AdS_6$ background has been found in \cite{Lozano:2012au}.
The field theory implications of the latter have yet to be exposed.

The superconformal index has emerged in recent years as a useful tool for studying superconformal field theories \cite{Kinney:2005ej}.
This is the superconformal analog of the Witten index, and is defined for a $d$-dimensional SCFT on $S^{d-1}\times \mbox{time}$
by
\begin{equation}
{\cal I} = \mbox{Tr}[(-1)^F\, e^{\mu_i q_i}] \,,
\end{equation}
where $q_i$ are charges that commute with a chosen supercharge.
This receives contributions only from states that are invariant under the chosen supercharge,
and therefore provides information on the number of such states and their charges.
This makes the index a useful diagnostic tool for revealing hidden structures in SCFT's such as enhanced global symmetries.
Furthermore,
since the index is a property of short (BPS) multiplets, it is protected from corrections due to continuous
deformations of the theory that preserve the chosen supercharge.
This makes it useful for testing various dualities, and for studying SCFT's that do not have a known Lagrangian,
but admit deformations into theories that do.

The study of superconformal indices in five dimensions was initiated in \cite{Kim:2012gu}. 
For a specific choice of the supercharge $Q$, the index of a 5d ${\cal N}=1$ SCFT is defined by
\begin{equation}
\mathcal{I}(x,y,{\bf z})={\rm Tr}\left[(-1)^F e^{-\beta\Delta} \,x^{2\,(j_1+R)}\,y^{2\,j_2}\,
{\bf z}^{\cal Q} \right] \,,
\end{equation}
where $\Delta = \{Q,S\} = E_0 - 2j_1 - 3R$, and $E_0$ is the conformal dimension, which is the energy
in the radially quantized theory.
The charges $j_1$, $j_2$ and $R$ are associated, respectively, to the Cartan $U(1)$'s of
the spatial $SU(2)_1\times SU(2)_2 \subset SO(5)$ and the $SU(2)_R$ R-symmetry.
The fugacities corresponding to the combinations that commute with $Q$ are denoted by $x$ and $y$.
Other commuting charges are denoted collectively by ${\cal Q}$, and their corresponding
fugacities by ${\bf z}$. This counts the $1/8$ BPS operators,  
for which $\Delta = 0$, namely $E_0 = 3R +2j_1$.

Using localization,
the 5d index was shown to admit a representation as an integral over the gauge group of the product of a perturbative (one-loop) component and an instanton component. In the following we will denote by $q$ the instanton fugacities, while we will reserve $\textbf{z}$ for other global commuting flavor-like fugacities. Then, the index can be written as
\begin{equation}
\label{generalindex}
\mathcal{I}(x,y,q,\mathbf{z})=\int [\mathcal{D}\alpha]\,I_{\rm inst}(x,y,q,\mathbf{z},\alpha)\,I_{\rm pert}(x,y,\mathbf{z},\alpha) \,.
\end{equation}
The integral over the gauge group is represented by an integral over the holonomy matrix $\alpha$ with the
appropriate Haar measure. The perturbative component is given by (a plethystic exponential) 
\begin{equation}
I_{\rm pert}(x,y,\mathbf{z},\alpha) = {\rm PE}[f(x,y,\mathbf{z},\alpha)]=e^{\sum_{n=1}^{\infty}\,\frac{f(x^n, y^n,\mathbf{z}^n, n\alpha)}{n}}\,,
\end{equation}
where $f(x,y,\mathbf{z},\alpha)$ is the sum of the single letter indices for the given theory.
The instanton component $I_{\rm inst}(x,y,q,\mathbf{z},\alpha)$ is given by a product of contributions of instantons located at the south pole of the $S^4$ and anti-instantons 
located at the north pole, 
and is related to  
Nekrasov's 4d  
instanton partition function \cite{Nekrasov:2002qd} (see \cite{Rodriguez-Gomez:2013dpa} for a recent discussion).
As its main application, the index was used in \cite{Kim:2012gu} to exhibit the enhanced $E_{N_f+1}$ global
symmetry of the $SU(2)$ theory with $N_f\leq 5$ flavors \cite{Seiberg:1996bd}.\footnote{For $N_f = 6,7$ 
some of the $E_{N_f + 1}$ currents come from the two-instanton sector, where there are some technical subtleties.
There is also a problem in generalizing to $USp(2N)$ with an antisymmetric hypermultiplet,
which is also expected to exhibit an enhanced global symmetry, related to the contribution 
of the antisymmetric field to the Nekrasov partition function.}

In this paper we begin the exploration of the superconformal indices of the 5d quiver theories
introduced in \cite{Bergman:2012kr}.
Since these theories possess gravity duals it is most interesting to study the large $N$ limit
of the index, and compare with the expectation from the gravity picture.
In this paper we will consider only the zero-instanton contribution to the index, namely
the perturbative superconformal index,
\begin{equation}
\mathcal{I}_{\rm pert}(x,y,\mathbf{z})=\int\,[\mathcal{D}\alpha]\, I_{pert}(x,y,\mathbf{z},\alpha) \,, 
\end{equation}
although we will also comment in the end on a possible large $N$ simplification of the instanton
contribution. Specifically, we will propose a general formula, motivated by AdS/CFT, for the perturbative large $N$ superconformal index of the 5d quiver SCFT's.

In the remainder of the paper we will drop the subscript ``pert",
and loosely refer to the perturbative index as the index.

\medskip

The outline for the rest of the paper is as follows.
We will begin in section \ref{USpsection} with the parent $USp(2N)$ theory,
and then consider the orbifold theories in section \ref{Orbifold_section}. Our proposal for the index of the orbifold theories appears in section \ref{proposal_section}.
In sections \ref{VS_section}, \ref{NVS_section} and \ref{Z3_section} we will perform
consistency checks for this proposal by explicitly computing the index for the $\mathbb{Z}_2$
and $\mathbb{Z}_3$ theories.
In particular, we will see that the large $N$ indices for the two different $\mathbb{Z}_2$
orbifold theories are the same, in agreement with the dual supergravity prediction. We will end in section \ref{conclusions} with some conclusions and future prospects.

\section{The $USp(2N)$ theory}
\label{USpsection}

The simplest class of 5d ${\cal N}=1$ fixed point theories with a known supergravity dual have a $USp(2N)$
gauge symmetry, an antisymmetric hypermultiplet, and $N_f$ fundamental hypermultiplets with $N_f\leq 7$.
This generalizes the $SU(2)$ theory of \cite{Seiberg:1996bd}.
For simplicity, we will concentrate on the case $N_f=0$.
Since the antisymmetric representation of $USp(2N)$ is real, the antisymmetric hypermultiplet splits into two
half-hypermultiplets, which transform as a doublet under a global mesonic symmetry $SU(2)_M$. There is an additional global symmetry $U(1)_I$ associated with the instanton current of the gauge group
$*\mbox{Tr}(F\wedge F)$.
The R-symmetry $SU(2)_R$ acts in the usual way on the components of the vector multiplet and hypermultiplet.

The supergravity dual of this theory was found in \cite{Brandhuber:1999np},
by realizing the corresponding gauge theory on D4-branes in a Type I' string theory background
with orientifold 8-planes and D8-branes.
The dual geometry is a warped product of $AdS_6$ and half of an $S^4$, with a metric and a dilaton given by
\begin{equation}
\label{metric_dilaton}
ds^2 \propto \sin^{-\frac{1}{3}}\alpha
\,\left[ds^2_{AdS_6}+\frac{4}{9}\,L^2\big(d\alpha^2+\cos^2\alpha\,d{s}_{S^3}^2\big) \right] \, , \;\;
e^{\Phi} \propto \sin^{-\frac{5}{6}}\alpha \,.
\end{equation}
Here $\alpha$ is the ``polar" angle on the $S^4$, which in this case ranges from $0$ (the ``equator" $S^3$)
to $\pi/2$ (the pole).
The warp factor reduces the symmetry of the compact piece to that of the $S^3$, namely to
$SO(4) \sim SU(2)_R\times SU(2)_M$.
The $U(1)_I$ symmetry is dual to the RR 1-form, and the instantons are dual to D0-branes.
Note that both the curvature and the dilaton diverge at the boundary $\alpha=0$,
signaling a breakdown of the perturbative supergravity description.
Indeed, the fixed point theory exhibits an enhancement of the global symmetry
$SO(2N_f)\times U(1)_I \rightarrow E_{N_f +1}$, which is not seen at the classical supergravity level.
In the Type I' string construction this enhancement is understood in terms of D0-branes that become massless
at the location of the orientifold \cite{Polchinski:1995df,Matalliotakis:1997qe,Bergman:1997py},
but this has not been understood yet at the level of the dual supergravity description.

\subsection{superconformal index}

The perturbative superconformal index of the $USp(2N)$ theory is given by
\begin{equation}
\mathcal{I}_1=\int [\mathcal{D}\alpha]\,{\rm PE}[f_V+f_A]\,,
\end{equation}
where the subscript is there to remind us of the degree of the orbifold.
The vector multiplet contribution to the single-particle index is (${\bf x}\equiv x,y$)
\begin{equation}
\label{vector_contribution}
f_V=i_V({\bf x})\,\Big[ 2\sum_{i\ne j}^N\,\cos\alpha_i\cos\alpha_j+2\,\sum_i^N\,\cos 2\alpha_i+N\Big] \,,
\end{equation}
where $\alpha_i \in [0,2\pi]$ are the holonomies associated to the Cartan subgroup, and $i_V({\bf x})$ is the vector mulitplet index,

\begin{equation}
i_V({\bf x})=-\frac{x\,(y+y^{-1})}{(1-x\,y)\,(1-x\,y^{-1})} \,.
\end{equation}
The antisymmetric hypermultiplet contributes 
\begin{equation}
f_H=i_H({\bf x})\, (z+z^{-1})\,\Big[\,2\,\sum_{i\ne j}^N\,\cos\alpha_i\,\cos\alpha_j+N\,\Big] \,,
\end{equation}
where $z$ is the fugacity associated with $U(1)_M\subset SU(2)_M$, and $i_H({\bf x})$
is the one particle index of a half-hypermultiplet,
\begin{equation}
i_H({\bf x})=\frac{x}{(1-x\,y)\,(1-x\,y^{-1})}\,.
\end{equation}
The Haar measure for $USp(2N)$ is
\begin{equation}
\label{Haar_measure}
[\mathcal{D}\alpha]=\prod_i d\alpha_i \,e^{\sum^N\,\log\sin^2\alpha_i+
\frac{1}{2}\,\sum_{i\ne j}^N\log\sin^2\Big(\frac{\alpha_i-\alpha_j}{2}\Big)
+\frac{1}{2}\,\sum_{i\ne j}^N\log\sin^2\Big(\frac{\alpha_i+\alpha_j}{2}\Big)} \,.
\end{equation}
One can then express the index as a partition function for a matrix model
\begin{equation}
\mathcal{I}_1=\int \prod_i^N d\alpha_i \,e^{-S} \,,
\end{equation}
with
\begin{equation}
S=2\sum_{i,\,j,\,m}\,\frac{1-i_V({\bf x}^m)-i_M({\bf x}^m)}{m}\cos m\alpha_i\cos m\alpha_j
+ \sum_{i,\,m}\frac{1-i_V({\bf x}^m)+i_M({\bf x}^m)}{m}\cos 2m\alpha_i \,,
\end{equation}
where $m$ is summed from 1 to $\infty$, and where, for convenience, we have defined
$i_M({\bf x}^m)\equiv i_H({\bf x}^m)\,(z^m + z^{-m})$.

Since we are interested in the large $N$ limit of the index,
we introduce the eigenvalue density
\begin{equation}
\rho(\theta) = \frac{1}{N}\sum_i^N \delta(\theta-\alpha_i)
\end{equation}
and define
\begin{equation}
\rho_m=\int_{-\pi}^{\pi} d\theta \,\rho(\theta)\,\cos m \theta \;, \quad m=0,1,2,\ldots
\end{equation}
We normalize the density such that $\int_{-\pi}^{\pi} d\theta\, \rho(\theta) = 1$.
The action becomes
\begin{equation}
S=2N^2\,\sum_{m=1}^\infty\,\frac{1-i_V({\bf x}^m)-i_M({\bf x}^m)}{m}\,\rho_m^2
+ N\sum_{m=1}^\infty\frac{1-i_V({\bf x}^m)+i_M({\bf x}^m)}{m}\,\rho_{2m} \,.
\end{equation}
In the large $N$ limit $\rho_m$ become continuous variables,
and we can replace the integrals over $\alpha_i$ with integrals over $\rho_m$.
This amounts to performing a saddle point approximation for the original integral.
The action is minimized by
\begin{equation}
{\rho}_{2m+1}=0\qquad {\rho}_{2m}=-\frac{1}{2N}\,\frac{1-i_V({\bf x}^m)+i_M({\bf x}^m)}
{1-i_V({\bf x}^{2\,m})-i_M({\bf x}^{2\,m})} \,.
\end{equation}
Performing the Gaussian integrals over the fluctuations then gives
\footnote{The index is normalized by dividing by the volume of the gauge group $\int [\mathcal{D}\alpha]$.}
\begin{equation}
\label{I1}
\mathcal{I}_1=\frac{e^{\sum\frac{1}{4\,m}\,\Big( \frac{[1-i_V({\bf x}^m)+i_M({\bf x}^m)]^2}
{1-i_V({\bf x}^{2m})-i_M({\bf x}^{2m})}-1\Big)}}{\prod\sqrt{1-i_V({\bf x}^{m})-i_M({\bf x}^{m})}} \,.
\end{equation}

In fact we can express the final result as a Plethystic exponential
\begin{equation}
\mathcal{I}_1= \mbox{PE}[G_1]\,,
\end{equation}
where $G_1$ is given in eq. (\ref{G_1}) in the Appendix.
In a precise sense, we can think of $G_1$ as the one-particle gauge-invariant index.
By expanding in a power series in one of the fugacities, we can express the index
as a sum of contributions of gauge invariant operators of increasing charge associated with the given fugacity.
Expanding in $x$ we find (to quadratic order):
\begin{equation}
\mathcal{I}_1=1+[1]_z\,x+\Big(1+2[2]_z+[1]_y\,[1]_z\Big)\,x^2+ {\cal O}(x^3) \,,
\end{equation}
where $[n]_z$ denotes the $SU(2)_M$ character in the spin $n/2$ representation, and similarly for $[n]_y$.
We can identify the different terms in terms of gauge-invariant operators as follows.
Let us denote the two complex scalar fields in the antisymmetric hypermultiplet by $A_\alpha$.
In 4d ${\cal N}=1$ language, $(A_1,A_2)$ is the pair of chiral superfields that make up the hypermultiplet.
The pair $(A_1,A_2)$ then transforms as an $SU(2)_M$ doublet, and the pair
$(A_1,A^\dagger_2)$ transforms as an $SU(2)_R$ doublet.
The ${\cal O}(x)$ term in the index corresponds to the basic meson operators $\mbox{Tr}A_\alpha \equiv (A_\alpha)_{ab}J^{ab}$.
The ${\cal O}(x^2)$ term contains four contributions.
The $SU(2)_M$ singlet corresponds to the scalar component of the $U(1)_I$ current,
given by the gaugino bilinear $\mbox{Tr}(\bar{\lambda}\lambda)$.
One of the $SU(2)_M$ triplets corresponds to $\mbox{Tr}(A_\alpha A_\beta)$, which are
the scalar components of the $SU(2)_M$ currents,
and the other corresponds to the double-trace operators $\mbox{Tr}A_\alpha \mbox{Tr}A_\beta$.\footnote{Note that
the F-term condition on the chiral ring does not affect this result, since it simply relates $A_1 A_2 = A_2 A_1$ \cite{Bergman:2012qh}.}
All of these have $R=1$ and $j_1=j_2=0$.
The fourth contribution corresponds to operators of the form $\mbox{Tr}(\partial A_\alpha)$,
which have $R=1/2$, $j_1=1/2$ and $j_2=\pm 1/2$.

\section{The orbifold theories}
\label{Orbifold_section}

There are three classes of orbifold models that yield new 5d ${\cal N}=1$ fixed point theories \cite{Bergman:2012kr}.
These can be engineered by replacing the flat 4d space transverse to the D4-branes and along the O8-D8 system
with an orbifold $\mathbb{C}^2/\mathbb{Z}_n$.
Generically this reduces the isometry to $SU(2)\times U(1)$.
The resulting 5d theories are ${\cal N}=1$ quiver gauge theories with bi-fundamental and antisymmetric matter,
as well as $N_f$ fundamentals.
We again take $N_f=0$.
The three classes of theories and their global symmetries are shown in Table~\ref{quiver_theories}.

\begin{table}[h!]
\begin{center}
\begin{tabular}{|l|l|l|l|}
\hline
model  & gauge group & matter & global (non-R) symmetry \\ \hline
$\mathbb{Z}_{2k}^{(1)}$ & $USp\times SU^{k-1}\times USp$   &  $(\funda_i,\funda_{i+1})$ &
$U(1)_M\times U(1)_B^{k-1}\times U(1)_I^{k+1}$  \\[5pt]
$\mathbb{Z}_{2k}^{(2)}$  & $SU^k$   &  $(\funda_i,\funda_{i+1}) + \asymm_1 + \asymm_{k}$  &
$U(1)_M\times U(1)_B^k\times U(1)_I^k$ \\[5pt]
$\mathbb{Z}_{2k+1}$ & $USp\times SU^k$   &  $(\funda_i,\funda_{i+1}) + \asymm_{k+1}$  &
$U(1)_M\times U(1)_B^k\times U(1)_I^{k+1}$\\
\hline
\end{tabular}
 \end{center}
\caption{5d orbifold quiver gauge theories. The groups are $USp(2N)$ and $SU(2N)$.}
\label{quiver_theories}
\end{table}

The two classes of even orbifold theories $\mathbb{Z}_{2k}^{(1)}$ and $\mathbb{Z}_{2k}^{(2)}$
are associated with a discrete choice in the action
of world-sheet parity on the twisted sector of the orbifold \cite{Polchinski:1996ry}.
The corresponding closed string backgrounds are known as the orbifold with and without vector structure, respectively.
The orbifold theories generically have three types of global (non-R) symmetries:
a single overall mesonic matter symmetry $U(1)_M$,
identified with the $U(1)$ part of the isometry of $\mathbb{C}^2/\mathbb{Z}_n$,
a number of baryonic matter $U(1)_B$ symmetries,
and a number of topological (instantonic) $U(1)_I$ symmetries.
The number of the latter is of course the number of gauge group factors.
In the $\mathbb{Z}_2$ theories the mesonic symmetry is enhanced to $SU(2)_M$
(as for the parent $USp(2N)$ theory).
In the $\mathbb{Z}_2^{(1)}$ theory there is a bi-fundamental hypermultiplet,
which, since this is a real representation of the gauge group, splits into two half-hypermultiplets
forming an $SU(2)_M$ doublet.
In the $\mathbb{Z}_2^{(2)}$ theory there are two $SU(2N)$-antisymmetric hypermultiplets,
which form an $SU(2)_M$ doublet. (In this case the gauge group representation is complex, so there is no splitting).
In either case, this is also clear from the string theory construction,
where it simply reflects the larger isometry of $\mathbb{C}^2/\mathbb{Z}_n$ for $n=2$.

The supergravity duals of these models are obtained by replacing $S^4$ with $S^4/\mathbb{Z}_n$,
where the $\mathbb{Z}_n$ acts freely on the $S^3$ base in (\ref{metric_dilaton}),
resulting in the lens space $S^3/\mathbb{Z}_n$.
The symmetry is therefore reduced to $SU(2)\times U(1)$ (except for the case of $n=2$),
which correspond respectively to $SU(2)_R$ and $U(1)_M$ in the field theory.
The full compact space $S^4/\mathbb{Z}_n$ has an $A_{n-1}$ fixed point singularity at the pole $\alpha = \pi/2$,
where one must include additional fields corresponding to the twisted sectors of the string theory.
In particular this includes a massless vector field for each twisted sector.
Taking into account the action of worldsheet parity leaves $k$ vectors for the odd orbifolds ($n=2\,k-1$)
and the even orbifolds ($n=2\,k$) with vector structure, and $k-1$ vectors for the even orbifolds without
vector structure. These correspond to the additional $U(1)_I$ symmetries beyond the one dual to the bulk RR 1-form.
They can also be described as reductions of the RR 3-form on the shrunk 2-cycles of the $A_{n-1}$ singularity.
In addition, the compact space has finite 2-cycles (dual to the shrunk ones), that similarly give rise to the additional
massless vectors dual to the baryonic symmetries $U(1)_B$.
In all, there are $n-1$ massless $U(1)$ vector multiplets coming from the 2-cycles, corresponding to the
baryonic and relative-instantonic symmetries, one from the RR 1-form corresponding to the overall instantonic symmetry,
and one from the isometry of $S^4/\mathbb{Z}_n$ (the latter is enhanced to $SU(2)$ for $n=1,2$).

Focusing on the two even orbifold models, we see that they are dual to the same metric-dilaton background,
with the same number of massless vector fields.
In fact they
are only distinguished by the presence of a trapped $B_2$ flux on the $k$'th 2-cycle of the $A_{2k-1}$ singularity
in the orbifold without vector structure \cite{Sen:1997pm}.

\subsection{orbifolding the index}
\label{proposal_section}

One could in principle compute the superconformal index of the orbifold theories directly from their field content,
but this becomes cumbersome for large $n$.
However the AdS/CFT correspondence suggests the following prescription \cite{Nakayama:2005mf, Gadde:2009dj}.
Starting with the $USp(2N)$ theory, we first want to project onto the operators that are invariant under the orbifold action.
Denoting by $\omega$ the generator of $\mathbb{Z}_n$, this amounts to replacing $z\rightarrow \omega^j z$
and summing over $j=0,\ldots n-1$.
This corresponds on the supergravity side to projecting the KK spectrum of the $AdS_6\times S^4$ background.
Then we need to add the contributions of additional operators, also invariant under $\mathbb{Z}_n$,
dual to the ``twisted sector" fields associated to the 2-cycles of the orbifold $S^4/\mathbb{Z}_n$.
Each cycle contributes a massless vector multiplet in $AdS_6$, and so we expect to see a contribution
of a dimension 3 BPS primary scalar operator from each such ``twisted sector"
(see for example \cite{D'Auria:2000ad,D'Auria:2000ay}).
This motivates our conjecture, that
\begin{equation}
\label{generaln}
\mathcal{I}_{n}={\rm PE}[G_n]
\qquad G_{n}=\frac{1}{n}\sum_{j=0}^{n-1}\,G_{1}(\omega^j\,z)+(n-1)\,\Delta \,,
\end{equation}
where $\Delta$ is the contribution corresponding to a dimension 3 BPS scalar and its derivatives, namely:
\begin{equation}
\label{Delta}
\Delta=\frac{x^2}{(1-xy)(1-xy^{-1})} \,.
\end{equation}
Expanding in $x$ for $n>2$ we find
\begin{equation}
\label{Zn_index}
\mathcal{I}_n=1+(n+1)\,x^2+ {\cal O}(x^3) \, .
\end{equation}
In particular, the contributions to $x^2$ in (\ref{Zn_index}) correspond to the $n+1$ conserved $U(1)$ currents
of the $\mathbb{Z}_n$ theory, including the $U(1)_M$, the $U(1)_B$'s and the $U(1)_I$'s.
For $n=2$ we get
\begin{equation}
\label{Z2_index_small_x}
\mathcal{I}_2=1+\left(2+[2]_z\right)\,x^2+ {\cal O}(x^3) \,,
\end{equation}
which exhibits the enhancement of $U(1)_M$ to $SU(2)_M$.
We will be more explicit about the operators corresponding to the $x^2$ contributions below.

Note that our conjecture implies that the superconformal index of the two even orbifold theories $\mathbb{Z}_{2k}^{(1)}$
and $\mathbb{Z}_{2k}^{(2)}$ is identical at large $N$.
This is what we expect from AdS/CFT.
Since the two supergravity backgrounds differ only by a discrete $B_2$ flux, the KK spectra should be virtually identical.

As consistency checks, we will now compute the superconformal indices of the first three orbifold theories directly.

\subsection{$\mathbb{Z}_2^{(1)}$ orbifold done explicitly}
\label{VS_section}

This theory has a $USp(2\,N)\times USp(2\,N)$ gauge symmetry and one bi-fundamental hypermultiplet.
The global symmetry is $SU(2)_M\times U(1)_I^2$.
We will associate the fugacity $z$ to the $U(1)_M\subset SU(2)_M$.
Each Haar measure and vector multiplet contribution is a copy of the $USp(2N)$ case (\ref{vector_contribution}), (\ref{Haar_measure}).
We will denote by $\alpha_i$ and $\beta_i$ the holonomies associated to the two $USp(2N)$ groups.
The contribution of the bi-fundamental hypermultiplet to the single-particle index is given by
\begin{equation}
f_H=4\, i_H({\bf x}) (z+z^{-1}) \sum_{i,\,j}^N\,\cos\alpha_i\,\cos\beta_j \,.
\end{equation}
Putting it all together, the matrix model action is given by
\begin{eqnarray}
S&=&\sum_{i,\,m}\frac{1-i_V({\bf x}^m)}{m}\,\cos2\,m\,\alpha_i+\sum_{i,\,m}\frac{1-i_V({\bf x}^m)}{m}\,\cos2\,m\,\beta_i  \nonumber\\
&+&  2\,\sum_{i,\,j,\,m}\frac{1-i_V({\bf x}^m)}{m}\,\cos m\,\alpha_i\,\cos m\,\alpha_j+ 2\,\sum_{i,\,j,\,m}\frac{1-i_V({\bf x}^m)}{m}\,\cos m\,\beta_i\,\cos m\,\beta_j
\nonumber\\
&-&  4\,\sum_{i,\,j,\,m}\frac{1-i_M({\bf x}^m)}{m}\,\cos m\,\alpha_i\,\cos m\,\beta_j \,,
\end{eqnarray}
where, as before $i_M \equiv (z+z^{-1})i_H$.
Introducing the two eigenvalue densities $\rho^{\alpha}$ and $\rho^{\beta}$ for the two gauge groups, one finds at large $N$:
\begin{eqnarray}
S&=&N \sum_{m}\frac{1-i_V(\bf{x}^m)}{m}\,\rho_{2m}^{\alpha}+N\sum_{m}\frac{1-i_V({\bf x}^m)}{m}\rho_{2m}^{\beta} +
2N^2\sum_{m}\frac{1-i_V({\bf x}^m)}{m}(\rho_{m}^{\alpha})^2
\nonumber\\
& + &  2N^2\,\sum_{m}\frac{1-i_V({\bf x}^m)}{m}(\rho_{m}^{\beta})^2
- 4N^2 \sum_{m}\frac{1-i_M({\bf x}^m)}{m}\rho_{m}^{\alpha}\rho_{m}^{\beta} \,.
\end{eqnarray}
The action is minimized by
\begin{equation}
{\rho}_{2\,m+1}^{\alpha}={\rho}_{2\,m+1}^{\beta}=0
\qquad {\rho}_{2\,m}^{\alpha}={\rho}_{2\,m}^{\beta}=-\frac{1}{2\,N}\,\frac{1-i_V(x^m)}{1-i_V(x^{2\,m}-i_M(x^{2\,m})} \,.
\end{equation}
Performing the Gaussian integrals then gives
\begin{equation}
\label{I2VS}
\mathcal{I}^{(1)}_{2}=\frac{e^{\sum\frac{1}{2\,m}\,\Big( \frac{[1-i_V({\bf x}^m)]^2}{1-i_V({\bf x}^{2m})-i_M({\bf x}^{2m})}-1\Big)}}
{\prod\sqrt{1-i_V({\bf x}^{m})-i_M({\bf x}^{m})}\,\sqrt{1-i_V({\bf x}^{m})+i_M({\bf x}^{m})}} \,.
\end{equation}
As in the case of the parent $USp(2N)$ theory, this result can be expressed as a plethystic exponential
\begin{equation}
\label{comp}
\mathcal{I}_{2}^{(1)}={\rm PE}[G_{2}] \,, 
\end{equation}
with $G_2$ given in eq.~(\ref{G_2}) in the Appendix.
It is straightforward to verify that the expression for $G_2$ agrees with (\ref{generaln}).

Let us now identify the gauge-invariant operators in this model corresponding to the ${\cal O}(x^2)$ contributions in (\ref{Z2_index_small_x}).
The two $SU(2)_M$ singlets are the scalar components of the
two $U(1)_I$ current multiplets, $\mbox{Tr}(\bar{\lambda}_1\lambda_1 \pm \bar{\lambda}_2\lambda_2)$,
where $\lambda_1, \lambda_2$ are the gauginos of the two $USp(2N)$ factors.
Denoting the two scalars of the bi-fundamental hypermultiplet as $X_\alpha$, 
the $SU(2)_M$ triplet contribution corresponds to
$\mbox{Tr}(X_\alpha X_\beta) \equiv (X_\alpha)^a_b (X_\beta)^c_d J_{ac} J^{bd}$,
which are the scalar components of the $SU(2)_M$ current multiplets.\footnote{As before, this result is not affected by
the F-term condition $X_1 J X_2 J= X_2 J X_1 J$ \cite{Bergman:2012qh}.}
Note that the quantum $\mathbb{Z}_2$ symmetry of the orbifold acts in the gauge theory by exchanging the
two gauge groups.
This shows that the operator $\mbox{Tr}(\bar{\lambda}_1\lambda_1 - \bar{\lambda}_2\lambda_2)$
is dual to a twisted sector state of the orbifold, whereas the others are dual to untwisted sector states.
Therefore we can associate the contribution of the former to (the ${\cal O}(x^2)$ term in) $\Delta$ (\ref{Delta}).

\subsection{$\mathbb{Z}_2^{(2)}$ orbifold done explicitly}
\label{NVS_section}

This theory has an $SU(2N)$ gauge symmetry and two antisymmetric hypermultiplets.
The global symmetry is $U(2)\times U(1)_I$, where the $U(2)$ is associated to the two complex matter multiplets.
This naturally splits into a baryonic $U(1)_B$ and a mesonic $SU(2)_M$.
We will denote the baryonic fugacity by $b$, and the mesonic $U(1)_M$ fugacity by $z$.

In computing the index, it is simpler to consider the closely related $U(2N)$ theory.
At large $N$ the indices are the same up to the contribution of the extra $U(1)$ vector multiplet.
In particular, the saddle point will coincide with that of the $SU(2N)$ theory for large $N$
(see for example \cite{Jafferis:2012iv}).
The Haar measure is given by
\begin{equation}
[\mathcal{D}\alpha]=\left[\prod_{i=1}^{2N} d\alpha_i\right] e^{\frac{1}{2}\,\sum_{i\ne j}^{2N}\,\log\sin^2\left(\frac{\alpha_i-\alpha_j}{2}\right)} \,.
\end{equation}
The vector multiplets contribute
\begin{equation}
f_V=i_V({\bf x}) \Big[\sum_{i,\,j}^{2N}\cos(\alpha_i-\alpha_j)-1\Big] \,,
\end{equation}
where we have subtracted the contribution of the $U(1)$ vector multiplet,
and the hypermultiplets contribute
\begin{equation}
f_H=i_M({\bf x}) \,b\,\sum_{i<j}^{2N}\,e^{i\,(\alpha_i+\alpha_j)}+i_M({\bf x})\,b^{-1}\,\sum_{i<j}^{2N}\,e^{-i\,(\alpha_i+\alpha_j)} \,.
\end{equation}
At large $N$ we introduce the eigenvalue density $\rho$ and define
\begin{equation}
\rho_m=\int d\theta\,\rho(\theta)\,\cos m\,\theta\qquad \sigma_m=\int d\theta\,\rho(\theta)\,\sin m\,\theta \,.
\end{equation}
After making a convenient change of variables,\footnote{The Jacobian of the transformation to $u_n, v_n$
is just a constant, so we can neglect it.}
\begin{equation}
u_n=(b^{\frac{n}{2}}+b^{-\frac{n}{2}})\,\sigma_n-i\,(b^{\frac{n}{2}}-b^{-\frac{n}{2}})\,\rho_n\qquad v_n
=(b^{\frac{n}{2}}+b^{-\frac{n}{2}})\,\rho_n+i\,(b^{\frac{n}{2}}-b^{-\frac{n}{2}})\,\sigma_n \,,
\end{equation}
we find
\begin{eqnarray}
\label{NVS_action}
S & = &N^2\sum_m\frac{1-i_V({\bf x}^m)+i_M({\bf x}^m)}{4m}\,u_m^2+N^2\sum_m\frac{1-i_V({\bf x}^m)+i_M({\bf x}^m)}{4m}\,v_m^2\nonumber \\
&+& N\sum_m\frac{i_M({\bf x}^m)}{2m}\,v_{2m}-\sum_m\frac{i_V({\bf x}^m)}{m} \,.
\end{eqnarray}
The action is minimized by
\begin{equation}
{u}_n={v}_{2\,n+1}=0\qquad {v}_{2\,n}=-\frac{2}{N}\,\frac{i_M(x^n)}{1-i_V(x^{2\,m}-i_M(x^{2\,m})} \,,
\end{equation}
and the integrals yield
\begin{equation}
\label{I2NVS}
\mathcal{I}^{(2)}_{2}=\frac{e^{\sum\left[\frac{1}{2m} \frac{(i_M({\bf x}^m))^2}{1-i_V({\bf x}^{2m})-i_M({\bf x}^{2m})}
-\frac{i_V({\bf x}^m)}{m}\right]}}{\prod\sqrt{1-i_V({\bf x}^{m})-i_M({\bf x}^{m})}\,\prod\sqrt{1-i_V({\bf x}^{m})+i_M({\bf x}^{m})}} \,.
\end{equation}
Using the explicit forms of $i_V({\bf x})$ and $i_M({\bf x})$, it is straightforward to show that
the numerators of (\ref{I2NVS}) and (\ref{I2VS}) are equal, and therefore that the large $N$ indices of the
two $\mathbb{Z}_2$ theories are identical, as they should be.
In other words the large $N$ index of this theory is also given by
\begin{equation}
\mathcal{I}_{2}^{(2)}={\rm PE}[G_{2}] \,, 
\end{equation}
in agreement with (\ref{generaln}).

Note that at large $N$ all the non-zero modes in (\ref{NVS_action}) are sharply peaked around 0,
so that $\rho(\theta) = 1/(2\pi) + {\cal O}(1/N)$.
This implies in particular that $\int_{-\pi}^{\pi} d\theta\,\theta\rho(\theta)=0$,
and therefore that $\sum_i^{2N}\alpha_i = 0$, as we should expect for the group $SU(2N)$.

Note also that the baryonic fugacity $b$ drops out from the index in the large $N$ limit.
This is consistent with the fact that gauge-invariant baryonic operators have
$\mathcal{O}(N)$ dimensions, and therefore contribute to the index with $\sim b\,x^{N}$.
Since $|x|<1$, at large $N$ their contribution vanishes.
This is also what would be seen in a dual supergravity computation of the index,
since baryons are dual to wrapped branes which become infinitely massive in the weak coupling limit.
In the large $N$ limit the superconformal index of the CFT includes only KK supergravity states.

The expansion in $x$ is of course the same as in the $\mathbb{Z}_2^{(1)}$ model (\ref{Z2_index_small_x}),
but the interpretation in terms of gauge-invariant operators will be different.
In this case one of the $SU(2)_M$ singlets is $\mbox{Tr}(\bar{\lambda}\lambda)$, the scalar component of the single
$U(1)_I$ current multiplet.
The other singlet and the triplet correspond to mesonic operators.
Let us denote by $A_\alpha$ the two complex scalars in the first antisymmetric hypermultiplet,
and by $A'_\alpha$ the two in the second.
In 4d ${\cal N}=1$ language, $(A_1,A_2)$ is the pair of chiral superfields that makes up the first
hypermultiplet, and likewise for the second.
The charge assignment of the fields follows the discussion in \cite{Bergman:2012qh}.
The pairs $(A_1, A_1')$ and $(A_2',A_2)$ transform as $SU(2)_M$ doublets,
and the pairs $(A_1,A_2^\dagger)$ and $(A'_1,A^{\prime\dagger}_2)$ transform as $SU(2)_R$ doublets.
In addition $A_1,A_1'$ carry baryon charge $B=+\frac{1}{2}$, and $A_2,A_2'$ have $B=-\frac{1}{2}$.
The four other operators contributing to the index at ${\cal O}(x^2)$, with $R=1$ and $j_1=j_2=0$, are therefore
the $SU(2)_M$ singlet $\mbox{Tr}(A_1 A_2 - A'_1A'_2)$, associated to the baryon current,
and the $SU(2)_M$ triplet $\{\mbox{Tr}(A_1 A'_2),\mbox{Tr}(A_1' A_2),\mbox{Tr}(A_1 A_2 + A'_1A'_2)\}$,
associated to the $SU(2)_M$ current.\footnote{Again, this is unaffected by the F-term conditions.
Note that these conditions were incorrectly stated in \cite{Bergman:2012qh}.
The correct 4d superpotential is $W = \mbox{Tr}(A_1 \Phi A_2 - A_1' \Phi A_2')$.
For $U(2N)$ this gives the F-term constraints $A_1 A_2 - A_1' A_2' =0$,
which appear to eliminate the $SU(2)_M$ singlet operator associated with the baryon current.
However for $SU(2N)$ the trace should be removed, namely
$A_1 A_2 - A_1' A_2' - \frac{1}{2N}\mbox{Tr}(A_1 A_2 - A_1' A_2') = 0$. This imposes no constraints on the above operators.}

In this case the quantum $\mathbb{Z}_2$ symmetry exchanges the two hypermultiplets (and multiplies
the vector multiplet by $-1$), so here we associate
the operator $\mbox{Tr}(A_1 A_2 - A'_1A'_2)$ to the twisted sector contribution $\Delta$.

\subsection{$\mathbb{Z}_3$ orbifold}
\label{Z3_section}

As a final consistency check of our formula for the large $N$ limit of the index of the orbifold CFT's (\ref{generaln}),
we now consider the $\mathbb{Z}_3$ orbifold.
This theory has a gauge symmetry $SU(2N)\times USp(2N)$, a hypermultiplet in the antisymmetric of $SU(2N)$,
and a bi-fundamental hypermultiplet.
We will use the convention that $i,j$ run from 1 to $N$, and $I,J$ run from 1 to $2N$.
As usual, the index is re-written as a matrix model. The Haar measure and vector contribution of the $USp(2N)$ piece together
give
\begin{equation}
S_{USp}= - \sum_{i,m} \frac{1-i_V({\bf x}^m)}{m}
\left[\cos 2m\alpha_i + 2\sum_j \cos m\alpha_i\cos m\alpha_j\right] \,.
\end{equation}
The analogous contributions of the $SU(2N)$ (or actually $U(2N)$) piece give
\begin{equation}
S_{SU}=-\sum_{I,J,m}\frac{1-i_V({\bf x}^m)}{m} \left[\cos m\beta_I\cos m\beta_J + \sin m\beta_I\sin m\beta_J\right]
-\sum_m\frac{i_V({\bf x}^m)}{m} \,.
\end{equation}
For brevity let us set both the mesonic and baryonic fugacities to one, $z=b=1$
(the baryonic fugacity will anyway drop out as before).
The bi-fundamental hypermultiplet contributes
\begin{equation}
S_{Bif}=\sum_{i,J,m}\,\frac{4\,i_H({\bf x}^m)}{m}\,\cos\alpha_i\,\cos\beta_J \,,
\end{equation}
and the $SU(2N)$-antisymmetric hypermultiplet contributes 
\begin{equation}
S_{Anti}=\sum_{I,J,m}\frac{i_H({\bf x}^m)}{m} \left[\cos m\beta_I\cos m\beta_J - \sin m\beta_I\sin m\beta_J \right]
- \sum_{I,m}\frac{i_H({\bf x}^m)}{m}\cos 2m\beta_I \,.
\end{equation}

Introducing the appropriate eigenvalue densities and taking the continuum large $N$ limit, the total action becomes
\begin{equation}
S = S_\rho + S_\sigma  -\sum_m\frac{i_V({\bf x}^m)}{m} \,,
\end{equation}
where
\begin{eqnarray}
S_{\sigma}= \mbox{} - 4N^2\sum_{m}\frac{1-i_V({\bf x}^m)+i_H({\bf x}^m)}{m}\,(\sigma_m^{\beta})^2
\end{eqnarray}
and
\begin{eqnarray}
S_{\rho} &=&\mbox{} -N\sum_{m} \frac{1-i_V({\bf x}^m)}{m}\,\rho^{\alpha}_{2m}
-2N\sum_m\frac{i_H({\bf x}^m)}{m}\,\rho^{\beta}_{2m}
- 2N^2\sum_{m}\frac{1-i_V({\bf x}^m)}{m}\,(\rho_m^{\alpha})^2\nonumber \\
 &-& 4N^2\sum_{m}\frac{1-i_V({\bf x}^m)-i_H({\bf x}^m)}{m}\,(\rho_m^{\beta})^2
 + 2N^2\sum_{m}\frac{4i_H({\bf x}^m)}{m}\,\rho_m^{\alpha}\,\rho_m^{\beta} \,.
\end{eqnarray}
The contribution of $S_\sigma$ to the index is simple to evaluate, and one finds
\begin{equation}
\mathcal{I}_{\sigma}= \int d\sigma^{\beta}\,e^{S_{\sigma}} =
\frac{1}{\prod\sqrt{1-i_V({\bf x}^m)+i_H({\bf x}^m)}} \,.
\end{equation}
To compute the contribution of $S_\rho$ it is convenient to first separate the odd and even modes of $\rho^{\alpha,\beta}$,
and then perform the integrals.
The final result is
\begin{eqnarray}
\label{I_3}
\mathcal{I}_3&=&e^{-\sum\frac{i_V({\bf x}^m)}{m}}\,\mathcal{I}_{\sigma}\,\mathcal{I_{\rho}} \\
& = &
\frac{{\rm PE}[f({\bf x})-i_V({\bf x})]}
{\prod\sqrt{1-i_V({\bf x}^m)+i_H({\bf x}^m)}\,
\prod_m\sqrt{ [1-i_V({\bf x}^m)]\,[1-i_V({\bf x}^m)-i_H({\bf x}^m)] -2\,i_H({\bf x}^m)^2}} \nonumber
\end{eqnarray}
where
\begin{equation}
f({\bf x})=\frac{[1-i_V({\bf x})]^2\,[1-i_V({\bf x}^2)-i_H({\bf x}^2)]
+ 2[1-i_V({\bf x}^2)]i_H({\bf x})^2
+ 4[1-i_V({\bf x})]i_H({\bf x})i_H({\bf x}^2)}
{4\Big[ [1-i_V({\bf x}^2)][1-i_V({\bf x}^2)-i_H({\bf x}^2)]-2i_H({\bf x}^2)^2\Big]} .
\end{equation}
Although this is certainly a rather cumbersome expression, one can check that it can be expressed as
\begin{equation}
{\cal I}_3 = {\rm PE}[G_3] \,,
\end{equation}
with $G_3$ given in eq.~(\ref{G_3}) (with $z=1$) in the Appendix. One can also show that that expression for $G_3$ also agrees with
with the $n=3$ (and $z=1$) version of eq.~(\ref{generaln}), namely 
\begin{equation}
G_{3}=\frac{1}{3}\sum_{j=0}^{2}\,G_1(e^{2\pi ij/3})+2\Delta \,.
\end{equation}


\section{Conclusions}\label{conclusions}

In this paper we have proposed a general formula (\ref{generaln}) for the large $N$ perturbative
superconformal index of the 5d ${\cal N}=1$ quiver gauge theories introduced in \cite{Bergman:2012kr}.
In very much the same spirit as in the case of orbifolds of 4d $\mathcal{N}=4$ SYM
\cite{Nakayama:2005mf, Gadde:2009dj}, the index for the 5d orbifold theories is produced by projecting
the parent theory to the invariant sector, and adding the contribution of the twisted sectors.
In particular, this procedure gives the same large $N$ index for the two classes of even orbifolds.
This agrees with the expectation from the supergravity duals,
which have identical geometries, and differ only in the presence of a trapped flux of the NSNS field
$B_2$ through the cycle corresponding to the middle twisted sector.
One still needs to perform the complete KK analysis in supergravity in order to compare with
the full index, but the consistency of our general formula (\ref{generaln}) with regards
to the twisted sector contributions already serves to reinforce the AdS/CFT dualities
proposed in \cite{Bergman:2012kr}.

The next natural step in this investigation is to compute the instanton contributions.
The authors of \cite{Kim:2012gu} did this to some extent for the parent $USp(2N)$ theory,
by including the 1-instanton contribution. This led to a beautiful verification of the
enhancement of the global symmtery to $E_{N_f +1}$ for $USp(2) = SU(2)$. 
The first question that comes to mind is whether this enhancement extends to $USp(2N)$ and to the quiver theories as well.
The Type I' brane construction suggests an enhancement, but it would be interesting to see
it at the level of the index. This requires including the instanton contributions to the index
of these theories. The results of  \cite{Kim:2012gu} suggest that at large $N$ there may
be a simplification. The large $N$ limit of their result for the $USp(2N)$ 1-instanton index is
\begin{equation}
\label{1-inst-large-N}
I_{\rm inst}^{k=1} \xrightarrow{N\rightarrow\infty}
\frac{x^2}{(1-xz)(1- xz^{-1})\,(1-xy)\,(1-xy^{-1})} \,.
\end{equation}
where $z$ is the fugacity corresponding to the global $SU(2)_M$ symmetry. This suggests that the full large $N$ instaton contribution, \textit{i.e.} $I_{\rm inst}(x,y,q,\mathbf{z},\alpha)$ in eq. (\ref{generalindex}), is given by
${\rm PE}[(q+q^{-1})I_{\rm inst}^{k=1}]$, 
which would manifestly exhibit symmetry enhancement since $q$ appears through $SU(2)$ characters.

It would also be very interesting to identify the supergravity duals of the instanton states
responsible for the symmetry enhancement.
Since in the Type I' brane construction these states correspond to D0-branes located
on the orientifold plane, it is natural to propose that in the near-horizon background
they correspond to D0-branes located at the boundary $\alpha = 0$ of the half-$S^4$.
Although the background is singular there, it is conceivable that some states are well-behaved.
This was the case, for example, for the dual giant gravitons on the Higgs branch
\cite{Bergman:2012qh}.
A D0-brane moving in $AdS_6$ and on $S^4$
would naturally account for the denominator in (\ref{1-inst-large-N}), however the overall
$x^2$ factor seems mysterious. We hope to report on these questions in the near future.

\section*{Acknowledgements}

O.B. and G.Z. are supported in part by the Israel Science Foundation under grant no. 392/09, and the US-Israel Binational Science Foundation under grant no. 2008-072.
D.R-G. is supported by a Spanish Government Ramon y Cajal fellowship RyC-2011-07593, and acknowledges support from the Spanish Ministry of Science through the research grant FPA2009-07122 and Spanish Consolider-Ingenio 2010 Programme CPAN (CSD2007-00042).

\appendix
\section{Some explicit formulas}
The single particle index for a vector multiplet and a half-hypermultiplet are given respectively by
\begin{equation}
i_V(x,y)=-\frac{x\,(y+y^{-1})}{(1-x\,y)\,(1-x\,y^{-1})} \qquad i_H(x,y)=\frac{x}{(1-x\,y)\,(1-x\,y^{-1})} \,.
\end{equation}
For convenience we also define the single particle mesonic index as $i_M(x,y,z)\equiv i_H(x,y)(z + z^{-1})$,
where $z$ is the fugacity associated with the mesonic symmetry, which is present in all
the models we discuss. Using the identity
\begin{equation}
\label{identity}
\prod_n\,\frac{1}{(1-x^n)^s}=e^{\sum\frac{1}{m}\frac{s\,x^m}{1-x^m}}={\rm PE}\left[\frac{sx}{1-x}\right] \,,
\end{equation}
we can express various products appearing in the expressions for the superconformal indices
in terms of Plethystic exponentials. For example,
the denominator in eq.~(\ref{I1}) is
\begin{equation}
\frac{1}{\prod\sqrt{1-i_V({\bf x}^{m})-i_M({\bf x}^{m})}}
={\rm PE}\left[\frac{1}{2}\left( \frac{xz}{1-xz}+\frac{xz^{-1}}{1-xz^{-1}}
-\frac{xy}{1-xy}-\frac{xy^{-1}}{1-xy^{-1}}\right)\right] .
\end{equation}
Therefore we can express the entire index as a Plethystic exponential, ${\cal I}_1 = \mbox{PE}[G_1]$, with
\begin{eqnarray}
\label{G_1}
G_{1}&=&\frac{1}{4}\left(\frac{[1-i_V(x)+i_M(x)]^2}{1-i_V(x^{2})-i_M(x^{2})}-1\right)
+ \frac{1}{2}\left( \frac{xz}{1-xz}+\frac{xz^{-1}}{1-xz^{-1}}-\frac{xy}{1-xy}-\frac{xy^{-1}}{1-xy^{-1}}\right) \nonumber\\[5pt]
&=& \frac{x(z+z^{-1}) + x^3(y+y^{-1})}{(1-xz)(1-xz^{-1})(1-xy)(1-xy^{-1})} \,.
\end{eqnarray}
Doing the same for the denominators in eqs.~(\ref{I2VS}) 
and (\ref{I_3}) we get
\begin{equation}
\label{G_2}
G_2 = \frac{x^2(3+z^2+z^{-2}) + x^3(y+y^{-1}) - x^4(z^2+z^{-2}) + x^5(y+y^{-1}) + x^6}
{(1-xy)(1-xy^{-1})(1-x^2z^2)(1-x^2z^{-2})} \,,
\end{equation}
and
\begin{equation}
\label{G_3}
G_3 = \frac{4x^2 + x^3(y+y^{-1}+z^3+z^{-3}) + x^4 + x^5(y+y^{-1}-2z^3-2z^{-3}) + x^7(y+y^{-1}) + x^8}
{(1-xy)(1-xy^{-1})(1-x^3z^3)(1-x^3z^{-3})} .
\end{equation}

\end{document}